\documentclass[
 reprint,
 amsmath,amssymb,
 aps,
 twocolumn,
 floatfix
]{revtex4-1}
\usepackage{natbib}
\usepackage[utf8x]{inputenc} 
\usepackage{graphicx}
\usepackage{amsmath}
\usepackage{textcomp}
\usepackage{cleveref}
\usepackage{float}

\begin{document}
\title{ \huge
A six-octave optical frequency comb \\ from a scalable few-cycle erbium fiber laser 
}
\author{Daniel M. B. Lesko,$^{1,2,\dagger,*}$ Henry Timmers,$^{1,\dagger}$ Sida Xing,$^{1,3}$ Abijith Kowligy,$^{1,3}$ Alexander J. Lind,$^{1,3}$ and Scott A. Diddams$^{1,3,*}$ \\
$^1$ Time and Frequency Division, National Institute of Standards and Technology, 325 Broadway, Boulder, Colorado 80305, USA \\
$^2$ Department of Chemistry, University of Colorado, 215 UCB, Boulder, Colorado 80309, USA\\
$^3$ Department of Physics, University of Colorado, 2000 Colorado Ave., Boulder, Colorado 80309, USA\\
$^\dagger$ These authors contributed equally to this work\\
$^*$ Corresponding author: Daniel Lesko (Daniel.Lesko@nist.gov) and Scott Diddams (Scott.Diddams@nist.gov)
}

\maketitle

\textbf{
A compact and robust coherent laser light source that provides spectral coverage from the ultraviolet to infrared is desirable for numerous applications, including heterodyne super resolution imaging\cite{yang_far-field_2016}, broadband infrared microscopy\cite{wetzel_imaging_1999}, protein structure determination\cite{williams_determination_1981}, and standoff atmospheric trace-gas  detection\cite{rieker_frequency-comb-based_2014}. 
Addressing these demanding measurement problems, laser frequency combs\cite{jones_carrier-envelope_2000} combine user-defined spectral resolution with sub-femtosecond timing and waveform control to enable new modalities of high-resolution, high-speed, and broadband spectroscopy\cite{coddington_dual-comb_2016,kowligy_infrared_2019,ideguchi_coherent_2013,bjork_direct_2016}. 
In this Letter we introduce a scalable source of near-single-cycle, 0.56~MW pulses generated from robust and low-noise erbium fiber (Er:fiber) technology, and we use it to generate a frequency comb that spans six octaves from the ultraviolet (350 nm) to mid-infrared (22500 nm). 
The high peak power allows us to exploit the second-order nonlinearities in infrared-transparent, nonlinear crystals (LiNbO$_3$, GaSe, and CSP) to provide a robust source of phase-stable infrared ultra-short pulses with simultaneous spectral brightness exceeding that of an infrared synchrotron\cite{bosch_computed_2000}.  
Additional cascaded second-order nonlinearities in LiNbO$_3$ lead to comb generation with four octaves of simultaneous coverage (0.350 to 5.6 $\mu$m).  
With a comb-tooth linewidth of 10 kHz at 193 THz, we realize a notable spectral resolving power exceeding $10^{10}$ across 0.86 PHz of bandwidth. 
We anticipate that this compact and accessible technology will open new opportunities for multi-band precision spectroscopy, coherent microscopy, ultra-high sensitivity nanoscopy, astronomical spectroscopy, and precision carrier-envelope phase (CEP) stable strong field phenomena.}

\begin{figure}[ht!]
  \centering
  \includegraphics[width=0.48\textwidth]{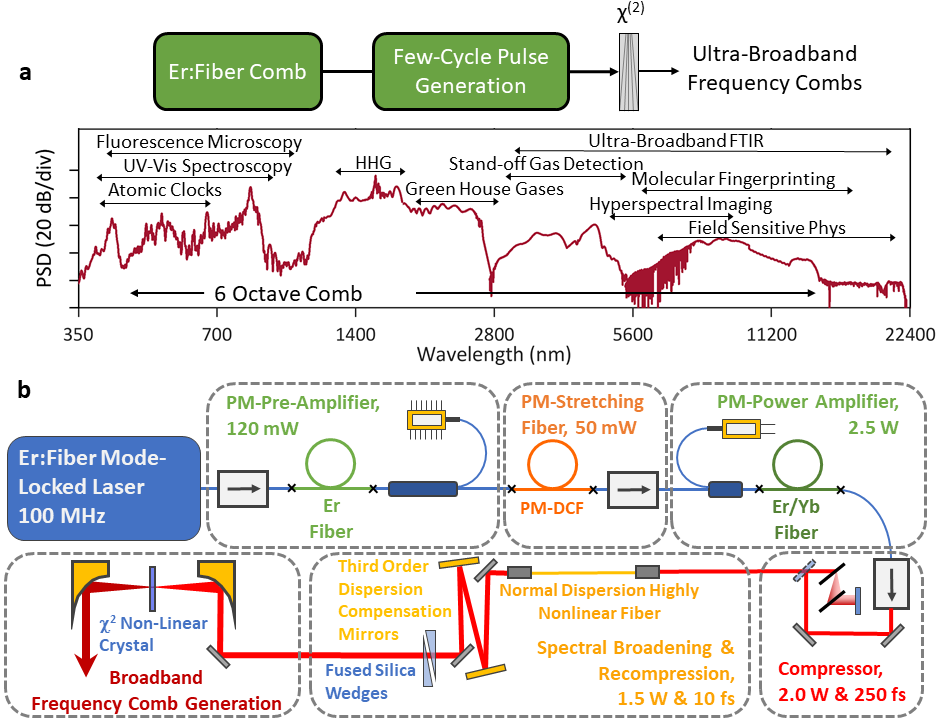}
\caption{Generation and applications of a six-octave frequency comb \textbf{a}, An erbium (Er) fiber mode-locked laser comb at 1550 nm seeds the generation of a few-cycle pulse to drive frequency conversion in second-order ($\chi^{(2)}$) nonlinear crystals. The resulting composite spectrum spans six octaves for a wide range of applications, as shown. \textbf{b}, Schematic of the fiber amplification and short pulse generation. See text for details. }
\end{figure}

Optical spectroscopy from the ultraviolet (UV) to the mid-infrared (MIR) has proven to be a critical technique for determining a molecule's precise structure and function in a non-destructive manner\cite{hollas_modern_nodate}.
Spectroscopy is generally accomplished using different light sources for each wavelength regime of interest.  
However, utilizing a single coherent source that can cover multiple absorption bands and spectroscopic regimes will enable correlated, high-fidelity measurements of simultaneous processes that are critical to many analytical science fields, from how water vibrates\cite{ramasesha_water_2013},
to how greenhouse gases bond to pollutant particles\cite{ostaszewski_effects_2018}, and how proteins in the skin are structured\cite{mendelsohn_determination_2006}.
The direct generation of broadband coherent sources is challenging, therefore nonlinear frequency conversion (i.e., $\chi^{(3)}$ supercontinuum\cite{dudley_supercontinuum_2006} and $\chi^{(2)}$ parametric\cite{vasilyev_multi-octave_2019} processes) has proven to be an indispensable technique for generating coherent light at otherwise inaccessible wavelengths.
The molecular fingerprint region represents one critical spectroscopic area in which the direct generation of a broadband, MIR source remains an open problem. 
Commercially available, multi-frequency quantum cascade lasers can typically only cover 1.8 THz simultaneously with $<$100 modes, so a promising alternative for broadband spectroscopy is to take inexpensive and robust near-infrared technology and transfer the stability to the MIR fingerprint region.
Leveraged by the global telecommunications infrastructure, erbium fiber (Er:fiber) technology offers commercially available ultra-low noise oscillators, versatile dispersion-engineered and nonlinear fibers, as well as an inexpensive, fiber-integrated off-the-shelf component catalogue.
This technology has demonstrated reliability, robustness, and utility for generating frequency combs in applications including optical clock comparisons\cite{udem_absolute_2001}, optical frequency division\cite{fortier_generation_2011}, generation of squeezed light states\cite{riek_direct_2015}, and high resolution spectroscopy\cite{coddington_dual-comb_2016}.

A simple way to generate ultra-broadband radiation is through intra-pulse difference frequency generation (IP-DFG), where an ultra-short pulse ($<$10 fs) is focused into a $\chi^{(2)}$ nonlinear medium where it undergoes nonlinear mixing and down-conversion\cite{timmers_molecular_2018}.
This process takes advantage of the large $\chi^{(2)}$ susceptibility of the medium as opposed to the much weaker $\chi^{(3)}$ nonlinearity.
An IP-DFG comb has a simple and robust architecture (no delay stage for temporal overlap, or cavity for enhancement) while generating extremely broadband and coherent radiation\cite{steinle_ultra-stable_2016,maidment_molecular_2016,gambetta_milliwatt-level_2013,sobon_high-power_2017}.
The down converted light is passively carrier-envelope stabilized and can be used as a seed for phase stable mid-infrared optical parametric chirped pulse amplification\cite{elu_high_2017} as well as variety of field sensitive processes including high harmonic generation\cite{pupeikis_water_2020}, laser-induced electron diffraction\cite{amini_imaging_2019}, and sensitive electro-optic sampling\cite{kowligy_infrared_2019,pupeza_field-resolved_2020}.

The challenge in generating an IP-DFG comb is two-fold: (i) the requirement for a few cycle driving pulse at high repetition rates ($\geq$100 MHz) and (ii) the low efficiency of the IP-DFG process itself. In previous papers\cite{timmers_molecular_2018,kowligy_infrared_2019,lind_mid-infrared_2020}, we presented a few-cycle Er:fiber comb generated from a core-pumped Er:fiber amplifier. The few-cycle comb was then used as a driver for IP-DFG to generate MIR light anywhere from 3~to~27 {\textmu}m. However, due to the limited power-scaling of the core-pumped Er:fiber, we were limited to  $<$1 mW of MIR comb light from the IP-DFG process.
In this Letter, we instead utilize a cladding pumped Erbium/Ytterbium co-doped fiber amplifier \cite{elahi_175_2017} (EYDFA) and demonstrate broadening and subsequent compression of this amplified pulse using a normal dispersion highly nonlinear fiber (ND-HNLF) to generate few cycle pulses at 1550 nm with up to 0.56 MW of peak power.
Using this high power, near-single-cycle comb source, we perform ultra-broadband frequency conversion in infrared $\chi^{(2)}$ nonlinear crystals to generate multi-octave combs spanning from the MIR to the UV.

\begin{figure}[ht!]
  \centering
  \includegraphics[width=0.47\textwidth]{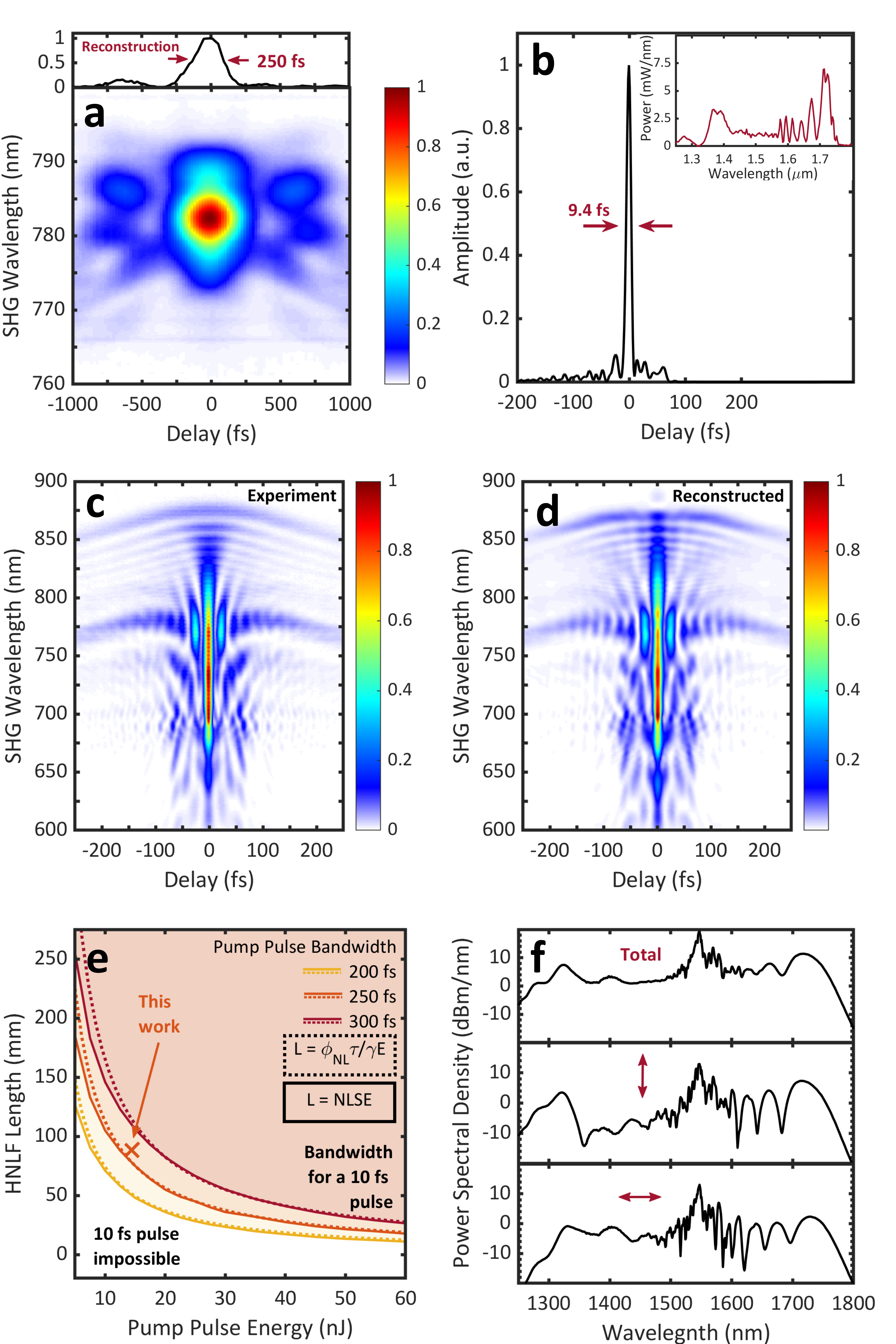}
\caption{Data and simulations demonstrating scalable near-single-cylce pulse generation. \textbf{a}, Second harmonic generation frequency resolved optical gating (SHG-FROG) of the chirped pulse amplifier output and the reconstructed 250~fs pulse. \textbf{b}, Retrieved temporal intensity profile of the 9.4 fs pulse and its spectrum (inset) that result from propagation in the ND-HNLF. \textbf{c}, Experimental SHG-FROG of the  near-single-cycle pulse pulse. \textbf{d}, Reconstructed SHG-FROG of the same with an RMS error of 1.5~\%. \textbf{e}, Nonlinear Schr\"{o}dinger equation (NLSE) simulations (solid lines) and analytic solution (dashed lines) for scaling HNLF length as a function of pump pulse energy and full width at half max (FWHM) bandwidth. Each contour represents the fiber length needed to support 10 fs pulses, given the specified pump pulse. \textbf{f}, Polarization dependent spectral broadening occurring in non-PM HNLF. The spectra measured in horizontal, vertical and both (total) polarizations are shown.}
\end{figure}

\begin{figure*}[ht!]
  \centering
  \includegraphics[width=0.90\textwidth]{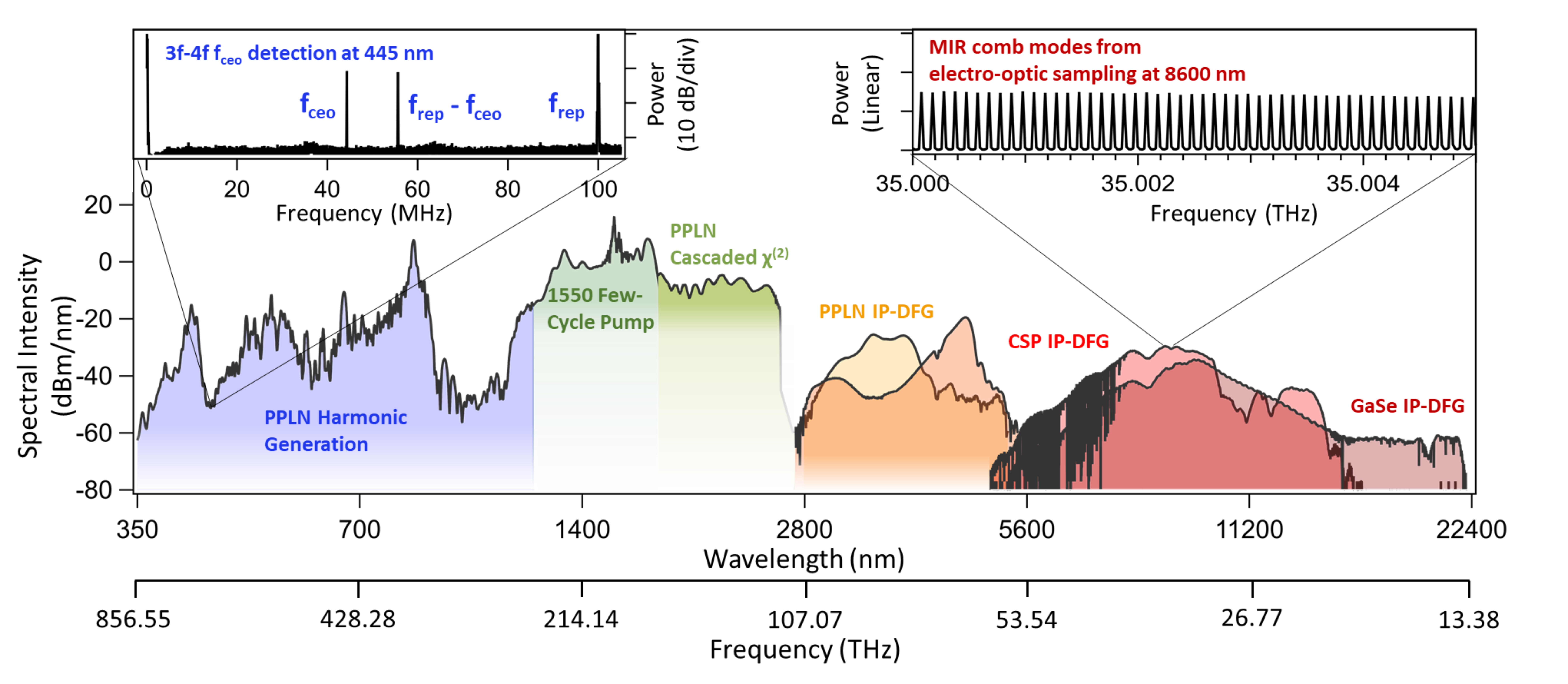}
\caption{Six-octave frequency comb. Frequency conversion from the 1550~nm 0.56~MW few-cycle pump is achieved via $\chi^{(2)}$ frequency conversion in PPLN, CSP and GaSe crystals. Shorter poling periods of the PPLN give simultaneous coverage from 350~to~5500~nm, while longer periods give selective coverage in the MIR. The broadest bandwidth spectra from CSP and GaSe are presented here, but different coverage can be achieved by angle-tuning the nonlinear crystals. \textbf{upper left inset}, f$_{\text{ceo}}$ measurement at 445 nm from harmonic generation in PPLN. \textbf{upper right inset}, Comb mode resolution of the infrared light from CSP measured with dual-comb electro-optic sampling.}
\end{figure*}

We describe our experimental design in Figure 1. The pulse first is amplified in a pre-amplifier and subsequently stretched with a dispersion compensating fiber before it seeds a chirped pulse EYDFA.  
The pre-amplifier limits the gain narrowing in the second stage of amplification.
The amplified pulse is subsequently compressed via a grating compressor to generate a 250 fs pump pulse with 2.5 W of average power, for which the reconstructed second harmonic generation frequency resolved optical gating (SHG-FROG) is shown in Figure 2a).
We then broaden the pump pulse in ND-HNLF to a bandwidth that supports a $<$10 fs pulse. Normal dispersion broadening minimizes modulation instability noise and ensures a clean phase accumulation that can be recompressed straightforwardly. Our near-single-cycle pulse is compressed using bulk UV fused silica and a pair of third order dispersion (TOD) compensating mirrors (Figure 2b), resulting in a compressed 9.4 fs, 0.56 MW pulse. 
The corresponding experimental and retrieved pulses (Figure 2c-d) show good agreement (1.5 \% RMS FROG error).
This spectral broadening process is modeled (Figure 2e) both by a simple analytic solution and using the nonlinear Schr\"{o}dinger equation (NLSE) for several amplifier regimes. Each contour represents the length of HNLF required to support a 10 fs pulse. The divergence of the dotted line from the solid occurs at HNLF fiber lengths for which the role of dispersion becomes important.
The non-PM HNLF also allows for polarization-dependent nonlinear effects, resulting in unequal amounts of self phase modulation (SPM) occurring along orthogonal axes (Figure 2f) as well as a reduction in the polarization extinction ratio (13 dB to 6 dB). This results in a longer HNLF required for the same level of spectral broadening (73~mm from theory to 140~mm in experiment).

Ultrabroadband $\chi^{(2)}$ frequency conversion is performed with this near-single-cycle pulse in three crystals to generate spectra that yield gap-less coverage over six octaves of bandwidth (Figure 3).
Periodically poled lithium niobate (PPLN) yields both 3.5~mW of MIR light (3~to~5~{\textmu}m) as well as harmonic generation and dispersive wave generation via cascaded $\chi^{(2)}$ (350~-~850~nm and 1.75~to~2.7~{\textmu}m respectively). To extend the wavelength further into the infrared, a 560 $\mu$m thick cadmium silicon phosphide (CSP) or a 1~mm thick gallium selenide (GaSe) are used to generate 3~mW and 1.6~mW respectively.
Three times greater power could be achieved with AR coatings on the crystals.
We confirm the resulting six octave source is a comb at both extremes. On the high frequency end, we observe an f$_{\text{ceo}}$ heterodyne beat at 445 nm, resulting from interference between the third and fourth harmonic (3f-4f heterodyne). And on the low frequency end, we directly resolve individual comb teeth through dual-comb electro-optic sampling\cite{kowligy_infrared_2019} with acquisition out to 22.4~{\textmu}m.
To switch between spectral ``modes" of the broadband comb, we only need to adjust the dispersion (fused silica wedges) to optimize for each spectral region.

 We compare our simple, fiber-based coherent frequency comb to the Advanced Light Source (ALS) infrared beamline operating conditions (500 mA) at Lawrence Berkeley National Labs and plot the resulting power spectral density (PSD) in Figure 4a\cite{bosch_computed_2000}.
 We can see that our IP-DFG comb outperforms the synchrotron across the MIR spectrum.  Such an experimental configuration not only realizes a table-top ``Er:fiber infrared synchrotron," but also provides a CEP-stable train of few cycle pulses. 
 The corresponding temporal waveforms generated through this process are nearly transform limited and few-cycle in duration (measured via electro-optic sampling, Figure 4b and d). The residual phase measured in Figure 4c and e arises from higher-order dispersion from a germanium beam splitter as well as atmospheric water contamination.
 This few cycle and near transform limited infrared pulse, enabled by extremely broad phase matching and precise pump dispersion control, is not possible in other comb systems\cite{steinle_ultra-stable_2016,smolski_half-watt_2018,chaitanya_kumar_high-power_2015,sobon_high-power_2017}.

\begin{figure}[ht!]
  \centering
  \includegraphics[width=.4\textwidth]{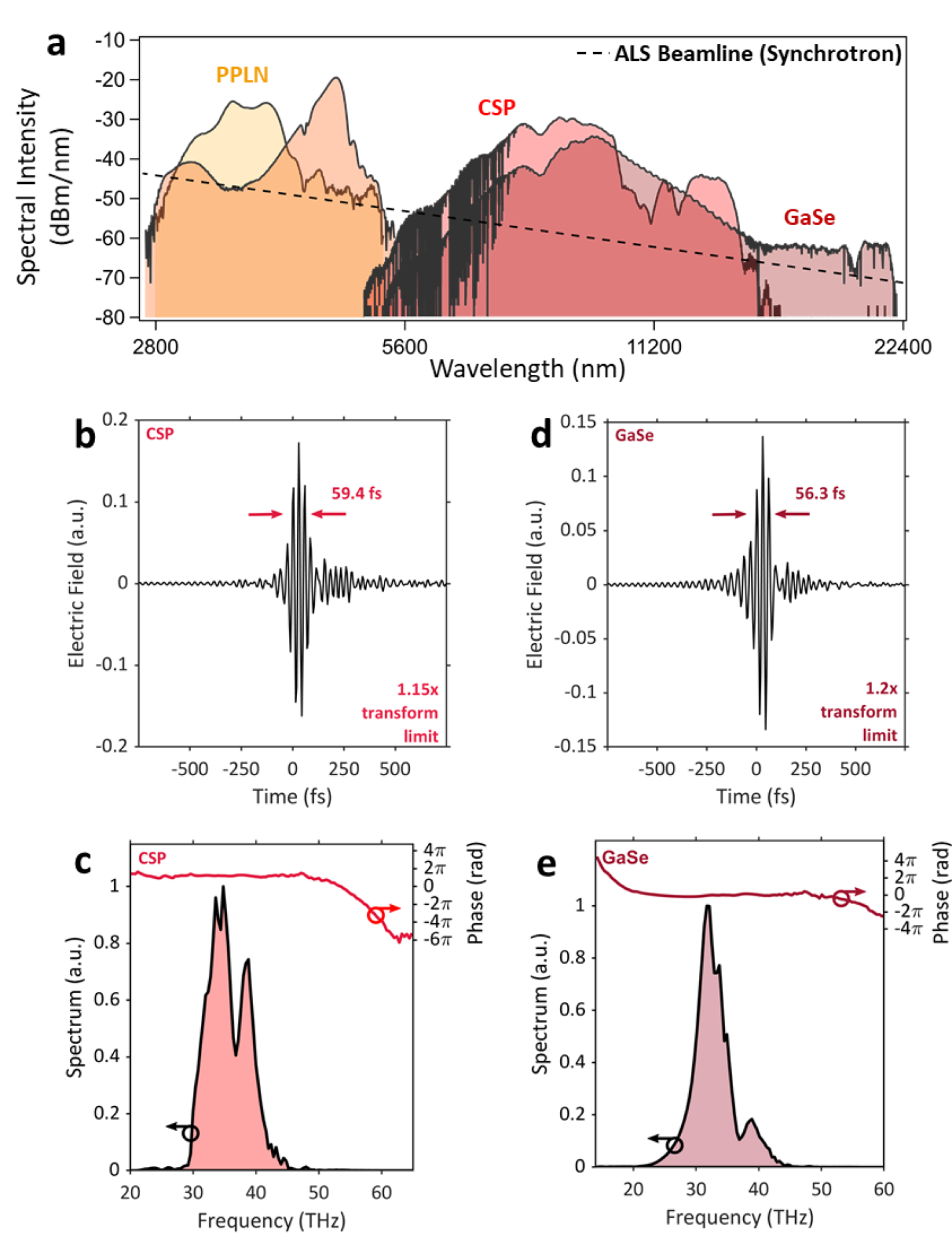}
\caption{MIR few-cycle pulses and frequency combs. \textbf{a}, Comparison of MIR comb generation from IP-DFG with the Advanced Light Source (ALS) beamline ($\sim$500~mA). \textbf{b} and \textbf{c}, Measured electric field from CSP and the resulting spectrum and phase. \textbf{d} and \textbf{e},Measured electric field from GaSe and the resulting spectrum and phase.}
\end{figure}

In summary, we present a simple, robust, and scalable architecture for generating megawatt scale peak power few-cycle pulses with ultra-low noise Er:fiber comb technology.
With these few-cycle pulses, we are able to generate a coherent ``light bulb", spanning six octaves (four simultaneously), utilizing only the second-order susceptibility of nonlinear crystals.
With this new turnkey and ultra-broadband coherent source, we will explore new spectroscopic modalities, perform sensitive precision microscopy and imaging, investigate coupled transitions in new quantum materials with nanoscopy, and determine mechanistic detail of chemical and biological processes over 6 octaves.
These results provide a clear path towards scaling few-cycle Er:fiber combs to $>$10 TW/cm$^2$ intensity regime, thus enabling the direct generation of extreme ultraviolet frequency combs and attosecond pulses without the need for complex enhancement cavity geometries enabling new regimes of physics to be explored.

\section*{Methods}
\textbf{Short pulse generation}\\
An ultra-low noise 100 MHz Menlo comb is amplified to 120~mW in an erbium doped fiber amplifier, and then stretched in PM dispersion compensation fiber and amplified to 2.5 W in an EYDFA. The pulse is compressed via a grating pair, and subsequently broadened in 14~cm of normal dispersion (D = -1.0 $\frac{\text{ps}}{\text{nm}{\cdot}\text{km}}$) HNLF. The pulse is then compressed in 5 cm bulk UV fused silica and third order compensation mirrors (SHG FROG Figures 2c-d).

\textbf{Spectral generation and measurement}\\
A 2 mm fan-out PPLN (26~to~35~{\textmu}m) is used to generate spectra from 350~to~5000 nm. Longer poling periods yield more targeted 3~to~5~$\mu$m light, while the shorter periods allow for harmonic generation (covering 350~to~850 nm with 30~mW total) as well as cascaded $\chi^{(2)}$ broadening (1.75~to~2.7~{\textmu}m).

A combination of optical spectrum analysers (350~to~1200, 700~to~1700, and 1200~to~2400~nm) were used to measure the visible and NIR spectra. The 1.8~to~5.5~$\mu$m section was measured with a commercial Fourier transform spectrometer (FTIR). The spectra from 5~to~22.4 {\textmu}m were measured using dual-comb electro-optic sampling\cite{kowligy_infrared_2019} to retrieve the electric field and spectrum. 
The 1.8~to~2.7~{\textmu}m dispersive wave was measured both on an OSA and the FTIR, but due to the smaller dynamic range of the FTIR and the higher noies floor of the OSA, the FTIR's data was scaled to match the calibrated PSD on the OSA.

\section*{Acknowledgements}
The mention of specific companies, products or tradenames does not constitute an endorsement by NIST.
The authors thank Thomas Schibli for his contributions and P. Schunemann and K. Zawilski at BAE for providing the CSP crystal, as well as I. Coddington, D. Carlson, and M. Hummon for their manuscript feedback.
DMBL and AK acknowledge award 70NANB18H006 from the National Institute of Standards and Technology (NIST). This research was supported by the Defense Advanced Research Projects Agency SCOUT Program, the Air Force Office of Scientific Research (FA9550-16-1-0016), and NIST.

\section*{Supplemental}
\begin{figure}[ht]
  \centering
  \includegraphics[width=.4\textwidth]{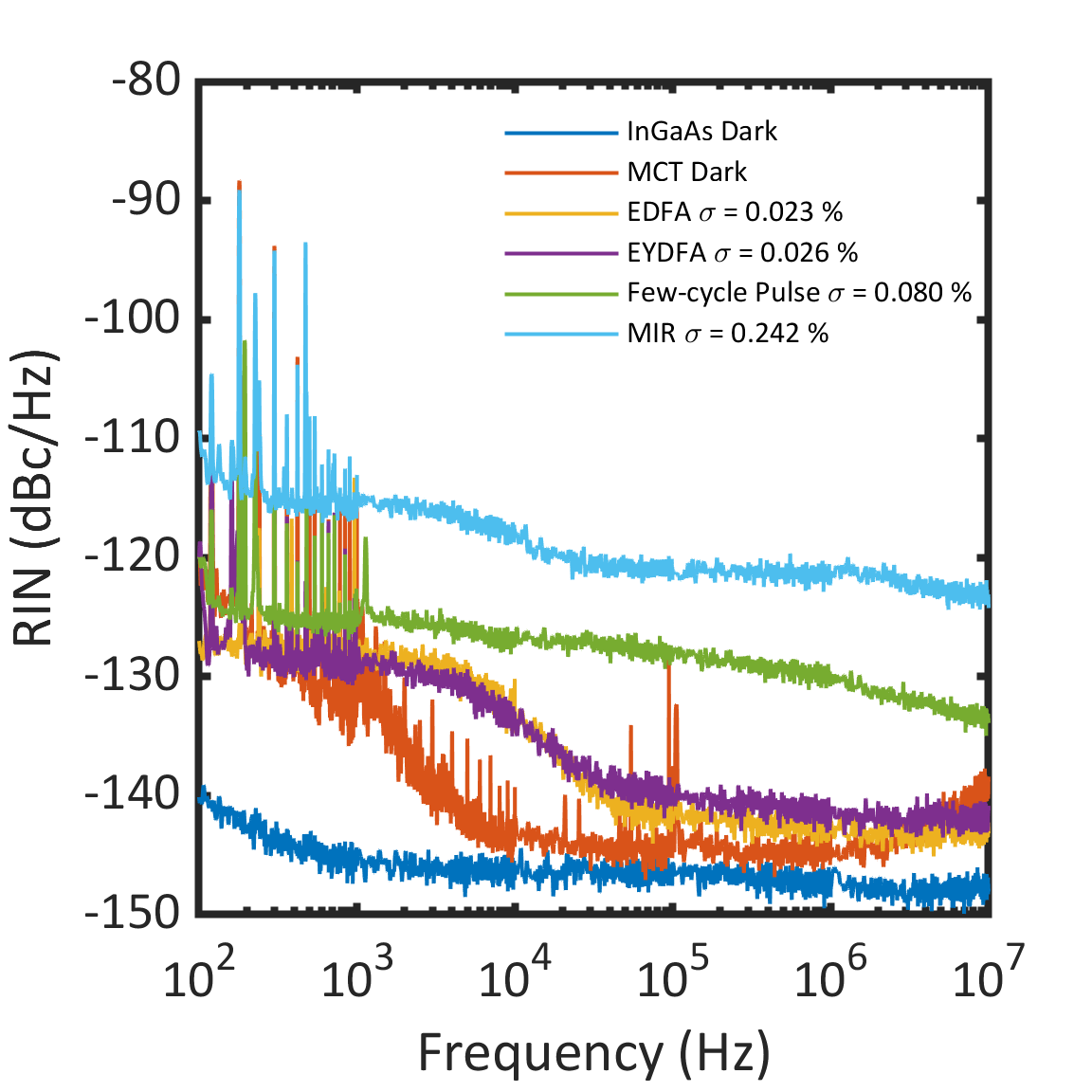}
\caption{Measured relative intensity noise of the system.}
\end{figure}
To characterize the intensity noise of the system, we measured the relative intensity noise (RIN) at each stage of the comb system. The measurements are summarized in Figure 1. All measurements in the NIR were done on an InGaAs detector (InGaAs Dark). We show that there is very little added noise from the EYDFA (0.026 \%) on the EDFA (0.023 \%). The oscillator has similar noise to the EDFA, as discussed in our previous paper\cite{timmers_molecular_2018}. The few cycle pulse has more noise (0.083 \%) but this can be mostly eliminated by utilizing a PM-HNLF due to other nonlinearities occurring in a non-PM fiber. The RIN of the MIR (0.242 \%) was measured on a fast liquid nitrogen cooled MCT detector (MCT Dark). The noise in the MIR is primarily due to other nonlinearities, besides SPM, occurring in the HNLF adding polarization noise.

\bibliography{MyLibrary.bib}

\begin{thebibliography}{10}
\expandafter\ifx\csname url\endcsname\relax
  \def\url#1{\texttt{#1}}\fi
\expandafter\ifx\csname urlprefix\endcsname\relax\def\urlprefix{URL }\fi
\providecommand{\bibinfo}[2]{#2}
\providecommand{\eprint}[2][]{\url{#2}}

\bibitem{yang_far-field_2016}
\bibinfo{author}{Yang, F.}, \bibinfo{author}{Tashchilina, A.},
  \bibinfo{author}{Moiseev, E.~S.}, \bibinfo{author}{Simon, C.} \&
  \bibinfo{author}{Lvovsky, A.~I.}
\newblock \bibinfo{title}{Far-field linear optical superresolution via
  heterodyne detection in a higher-order local oscillator mode}.
\newblock \emph{\bibinfo{journal}{Optica}} \textbf{\bibinfo{volume}{3}},
  \bibinfo{pages}{1148--1152} (\bibinfo{year}{2016}).

\bibitem{wetzel_imaging_1999}
\bibinfo{author}{Wetzel, D.~L.} \& \bibinfo{author}{LeVine, S.~M.}
\newblock \bibinfo{title}{Imaging {Molecular} {Chemistry} with {Infrared}
  {Microscopy}}.
\newblock \emph{\bibinfo{journal}{Science}} \textbf{\bibinfo{volume}{285}},
  \bibinfo{pages}{1224--1225} (\bibinfo{year}{1999}).

\bibitem{williams_determination_1981}
\bibinfo{author}{Williams, R.~W.} \& \bibinfo{author}{Dunker, A.~K.}
\newblock \bibinfo{title}{Determination of the secondary structure of proteins
  from the amide {I} band of the laser {Raman} spectrum}.
\newblock \emph{\bibinfo{journal}{Journal of Molecular Biology}}
  \textbf{\bibinfo{volume}{152}}, \bibinfo{pages}{783 -- 813}
  (\bibinfo{year}{1981}).

\bibitem{rieker_frequency-comb-based_2014}
\bibinfo{author}{Rieker, G.~B.} \emph{et~al.}
\newblock \bibinfo{title}{Frequency-comb-based remote sensing of greenhouse
  gases over kilometer air paths}.
\newblock \emph{\bibinfo{journal}{Optica}} \textbf{\bibinfo{volume}{1}},
  \bibinfo{pages}{290} (\bibinfo{year}{2014}).

\bibitem{jones_carrier-envelope_2000}
\bibinfo{author}{Jones, D.~J.}
\newblock \bibinfo{title}{Carrier-{Envelope} {Phase} {Control} of {Femtosecond}
  {Mode}-{Locked} {Lasers} and {Direct} {Optical} {Frequency} {Synthesis}}.
\newblock \emph{\bibinfo{journal}{Science}} \textbf{\bibinfo{volume}{288}},
  \bibinfo{pages}{635--639} (\bibinfo{year}{2000}).

\bibitem{coddington_dual-comb_2016}
\bibinfo{author}{Coddington, I.}, \bibinfo{author}{Newbury, N.} \&
  \bibinfo{author}{Swann, W.}
\newblock \bibinfo{title}{Dual-comb spectroscopy}.
\newblock \emph{\bibinfo{journal}{Optica}} \textbf{\bibinfo{volume}{3}},
  \bibinfo{pages}{414} (\bibinfo{year}{2016}).

\bibitem{kowligy_infrared_2019}
\bibinfo{author}{Kowligy, A.~S.} \emph{et~al.}
\newblock \bibinfo{title}{Infrared electric field sampled frequency comb
  spectroscopy}.
\newblock \emph{\bibinfo{journal}{Science Advances}}
  \textbf{\bibinfo{volume}{5}}, \bibinfo{pages}{eaaw8794}
  (\bibinfo{year}{2019}).

\bibitem{ideguchi_coherent_2013}
\bibinfo{author}{Ideguchi, T.} \emph{et~al.}
\newblock \bibinfo{title}{Coherent {Raman} spectro-imaging with laser frequency
  combs}.
\newblock \emph{\bibinfo{journal}{Nature}} \textbf{\bibinfo{volume}{502}},
  \bibinfo{pages}{355--358} (\bibinfo{year}{2013}).

\bibitem{bjork_direct_2016}
\bibinfo{author}{Bjork, B.~J.} \emph{et~al.}
\newblock \bibinfo{title}{Direct frequency comb measurement of {OD} + {CO} →
  {DOCO} kinetics}.
\newblock \emph{\bibinfo{journal}{Science}} \textbf{\bibinfo{volume}{354}},
  \bibinfo{pages}{444--448} (\bibinfo{year}{2016}).

\bibitem{bosch_computed_2000}
\bibinfo{author}{Bosch, R.~A.}
\newblock \bibinfo{title}{Computed flux and brightness of infrared edge and
  synchrotron radiation}.
\newblock \emph{\bibinfo{journal}{Nuclear Instruments and Methods in Physics
  Research. Section A, Accelerators, Spectrometers, Detectors and Associated
  Equipment}} \textbf{\bibinfo{volume}{454}}, \bibinfo{pages}{497--505}
  (\bibinfo{year}{2000}).

\bibitem{hollas_modern_nodate}
\bibinfo{author}{Hollas, J., Michael}.
\newblock \emph{\bibinfo{title}{Modern {Spectroscopy}}}
  (\bibinfo{publisher}{Wiley}), \bibinfo{edition}{4th} edn.

\bibitem{ramasesha_water_2013}
\bibinfo{author}{Ramasesha, K.}, \bibinfo{author}{De~Marco, L.},
  \bibinfo{author}{Mandal, A.} \& \bibinfo{author}{Tokmakoff, A.}
\newblock \bibinfo{title}{Water vibrations have strongly mixed intra- and
  intermolecular character}.
\newblock \emph{\bibinfo{journal}{Nature Chemistry}}
  \textbf{\bibinfo{volume}{5}}, \bibinfo{pages}{935--940}
  (\bibinfo{year}{2013}).

\bibitem{ostaszewski_effects_2018}
\bibinfo{author}{Ostaszewski, C.~J.} \emph{et~al.}
\newblock \bibinfo{title}{Effects of {Coadsorbed} {Water} on the
  {Heterogeneous} {Photochemistry} of {Nitrates} {Adsorbed} on {TiO}
  $_{\textrm{2}}$}.
\newblock \emph{\bibinfo{journal}{The Journal of Physical Chemistry A}}
  \textbf{\bibinfo{volume}{122}}, \bibinfo{pages}{6360--6371}
  (\bibinfo{year}{2018}).

\bibitem{mendelsohn_determination_2006}
\bibinfo{author}{Mendelsohn, R.}, \bibinfo{author}{Flach, C.~R.} \&
  \bibinfo{author}{Moore, D.~J.}
\newblock \bibinfo{title}{Determination of molecular conformation and
  permeation in skin via {IR} spectroscopy, microscopy, and imaging}.
\newblock \emph{\bibinfo{journal}{Biochimica et Biophysica Acta (BBA) -
  Biomembranes}} \textbf{\bibinfo{volume}{1758}}, \bibinfo{pages}{923 -- 933}
  (\bibinfo{year}{2006}).

\bibitem{dudley_supercontinuum_2006}
\bibinfo{author}{Dudley, J.~M.}, \bibinfo{author}{Genty, G.} \&
  \bibinfo{author}{Coen, S.}
\newblock \bibinfo{title}{Supercontinuum generation in photonic crystal fiber}.
\newblock \emph{\bibinfo{journal}{Reviews of Modern Physics}}
  \textbf{\bibinfo{volume}{78}}, \bibinfo{pages}{1135--1184}
  (\bibinfo{year}{2006}).

\bibitem{vasilyev_multi-octave_2019}
\bibinfo{author}{Vasilyev, S.} \emph{et~al.}
\newblock \bibinfo{title}{Multi-octave visible to long-wave {IR} femtosecond
  continuum generated in {Cr}:{ZnS}-{GaSe} tandem}.
\newblock \emph{\bibinfo{journal}{Optics Express}}
  \textbf{\bibinfo{volume}{27}}, \bibinfo{pages}{16405} (\bibinfo{year}{2019}).

\bibitem{udem_absolute_2001}
\bibinfo{author}{Udem, T.} \emph{et~al.}
\newblock \bibinfo{title}{Absolute {Frequency} {Measurements} of the {Hg} + and
  {Ca} {Optical} {Clock} {Transitions} with a {Femtosecond} {Laser}}.
\newblock \emph{\bibinfo{journal}{Physical Review Letters}}
  \textbf{\bibinfo{volume}{86}}, \bibinfo{pages}{4996--4999}
  (\bibinfo{year}{2001}).
\newblock \urlprefix\url{https://link.aps.org/doi/10.1103/PhysRevLett.86.4996}.

\bibitem{fortier_generation_2011}
\bibinfo{author}{Fortier, T.~M.} \emph{et~al.}
\newblock \bibinfo{title}{Generation of ultrastable microwaves via optical
  frequency division}.
\newblock \emph{\bibinfo{journal}{Nature Photonics}}
  \textbf{\bibinfo{volume}{5}}, \bibinfo{pages}{425--429}
  (\bibinfo{year}{2011}).

\bibitem{riek_direct_2015}
\bibinfo{author}{Riek, C.} \emph{et~al.}
\newblock \bibinfo{title}{Direct sampling of electric-field vacuum
  fluctuations}.
\newblock \emph{\bibinfo{journal}{Science}} \textbf{\bibinfo{volume}{350}},
  \bibinfo{pages}{420--423} (\bibinfo{year}{2015}).

\bibitem{timmers_molecular_2018}
\bibinfo{author}{Timmers, H.} \emph{et~al.}
\newblock \bibinfo{title}{Molecular fingerprinting with bright, broadband
  infrared frequency combs}.
\newblock \emph{\bibinfo{journal}{Optica}} \textbf{\bibinfo{volume}{5}},
  \bibinfo{pages}{727--732} (\bibinfo{year}{2018}).

\bibitem{steinle_ultra-stable_2016}
\bibinfo{author}{Steinle, T.}, \bibinfo{author}{Mörz, F.},
  \bibinfo{author}{Steinmann, A.} \& \bibinfo{author}{Giessen, H.}
\newblock \bibinfo{title}{Ultra-stable high average power femtosecond laser
  system tunable from 133 to 20 μm}.
\newblock \emph{\bibinfo{journal}{Optics Letters}}
  \textbf{\bibinfo{volume}{41}}, \bibinfo{pages}{4863} (\bibinfo{year}{2016}).

\bibitem{maidment_molecular_2016}
\bibinfo{author}{Maidment, L.}, \bibinfo{author}{Schunemann, P.~G.} \&
  \bibinfo{author}{Reid, D.~T.}
\newblock \bibinfo{title}{Molecular fingerprint-region spectroscopy from 5 to
  12 μm using an orientation-patterned gallium phosphide optical parametric
  oscillator}.
\newblock \emph{\bibinfo{journal}{Optics Letters}}
  \textbf{\bibinfo{volume}{41}}, \bibinfo{pages}{4261} (\bibinfo{year}{2016}).

\bibitem{gambetta_milliwatt-level_2013}
\bibinfo{author}{Gambetta, A.} \emph{et~al.}
\newblock \bibinfo{title}{Milliwatt-level frequency combs in the 8–14 μm
  range via difference frequency generation from an {Er}:fiber oscillator}.
\newblock \emph{\bibinfo{journal}{Optics Letters}}
  \textbf{\bibinfo{volume}{38}}, \bibinfo{pages}{1155} (\bibinfo{year}{2013}).

\bibitem{sobon_high-power_2017}
\bibinfo{author}{Soboń, G.}, \bibinfo{author}{Martynkien, T.},
  \bibinfo{author}{Mergo, P.}, \bibinfo{author}{Rutkowski, L.} \&
  \bibinfo{author}{Foltynowicz, A.}
\newblock \bibinfo{title}{High-power frequency comb source tunable from 2.7 to
  4.2 μm based on difference frequency generation pumped by an {Yb}-doped
  fiber laser}.
\newblock \emph{\bibinfo{journal}{Opt. Lett.}} \textbf{\bibinfo{volume}{42}},
  \bibinfo{pages}{1748--1751} (\bibinfo{year}{2017}).

\bibitem{elu_high_2017}
\bibinfo{author}{Elu, U.} \emph{et~al.}
\newblock \bibinfo{title}{High average power and single-cycle pulses from a
  mid-{IR} optical parametric chirped pulse amplifier}.
\newblock \emph{\bibinfo{journal}{Optica}} \textbf{\bibinfo{volume}{4}},
  \bibinfo{pages}{1024} (\bibinfo{year}{2017}).

\bibitem{pupeikis_water_2020}
\bibinfo{author}{Pupeikis, J.} \emph{et~al.}
\newblock \bibinfo{title}{Water window soft x-ray source enabled by a 25 {W}
  few-cycle 22 µm {OPCPA} at 100 {kHz}}.
\newblock \emph{\bibinfo{journal}{Optica}} \textbf{\bibinfo{volume}{7}},
  \bibinfo{pages}{168} (\bibinfo{year}{2020}).

\bibitem{amini_imaging_2019}
\bibinfo{author}{Amini, K.} \emph{et~al.}
\newblock \bibinfo{title}{Imaging the {Renner}–{Teller} effect using
  laser-induced electron diffraction}.
\newblock \emph{\bibinfo{journal}{Proceedings of the National Academy of
  Sciences}} \textbf{\bibinfo{volume}{116}}, \bibinfo{pages}{8173--8177}
  (\bibinfo{year}{2019}).
\newblock
  \urlprefix\url{http://www.pnas.org/lookup/doi/10.1073/pnas.1817465116}.

\bibitem{pupeza_field-resolved_2020}
\bibinfo{author}{Pupeza, I.} \emph{et~al.}
\newblock \bibinfo{title}{Field-resolved infrared spectroscopy of biological
  systems}.
\newblock \emph{\bibinfo{journal}{Nature}} \textbf{\bibinfo{volume}{577}},
  \bibinfo{pages}{52--59} (\bibinfo{year}{2020}).

\bibitem{lind_mid-infrared_2020}
\bibinfo{author}{Lind, A.~J.} \emph{et~al.}
\newblock \bibinfo{title}{Mid-{Infrared} {Frequency} {Comb} {Generation} and
  {Spectroscopy} with {Few}-{Cycle} {Pulses} and $\chi^{(2)}$ {Nonlinear}
  {Optics}}.
\newblock \emph{\bibinfo{journal}{Phys. Rev. Lett.}}
  \textbf{\bibinfo{volume}{124}}, \bibinfo{pages}{133904}
  (\bibinfo{year}{2020}).

\bibitem{elahi_175_2017}
\bibinfo{author}{Elahi, P.}, \bibinfo{author}{Kalaycıoğlu, H.},
  \bibinfo{author}{Li, H.}, \bibinfo{author}{Akçaalan, O.} \&
  \bibinfo{author}{Ilday, F.~O.}
\newblock \bibinfo{title}{175 fs-long pulses from a high-power single-mode
  {Er}-doped fiber laser at 1550 nm}.
\newblock \emph{\bibinfo{journal}{Optics Communications}}
  \textbf{\bibinfo{volume}{403}}, \bibinfo{pages}{381--384}
  (\bibinfo{year}{2017}).

\bibitem{smolski_half-watt_2018}
\bibinfo{author}{Smolski, V.} \emph{et~al.}
\newblock \bibinfo{title}{Half-{Watt} average power femtosecond source spanning
  3–8 µm based on subharmonic generation in {GaAs}}.
\newblock \emph{\bibinfo{journal}{Applied Physics B}}
  \textbf{\bibinfo{volume}{124}}, \bibinfo{pages}{101} (\bibinfo{year}{2018}).

\bibitem{chaitanya_kumar_high-power_2015}
\bibinfo{author}{Chaitanya~Kumar, S.} \emph{et~al.}
\newblock \bibinfo{title}{High-power femtosecond mid-infrared optical
  parametric oscillator at 7 μm based on {CdSiP}\_2}.
\newblock \emph{\bibinfo{journal}{Optics Letters}}
  \textbf{\bibinfo{volume}{40}}, \bibinfo{pages}{1398} (\bibinfo{year}{2015}).

\end{thebibliography}
\end{document}